\documentclass{sig-alternate-10pt}
\usepackage[margin=1in]{geometry}

\usepackage{graphicx}
\usepackage{balance}  
\usepackage{afterpage}
\usepackage{pifont}
\usepackage{algorithmic}
\usepackage{amsmath}
\usepackage{listings}
\usepackage{empheq}
\usepackage[framemethod=tikz,xcolor=true]{mdframed}

\usepackage{amsthm}
\usepackage{amsmath}
\usepackage{amsfonts}
\usepackage{colortbl}
\usepackage{paralist}
\usepackage{enumitem}
\usepackage{epsfig}
\usepackage{forest}
\usepackage{graphicx}
\usepackage{multicol}
\usepackage{multirow}
\usepackage{pgf}
\usepackage{pdflscape}
\usetikzlibrary{positioning}
\usepackage{rotating}
\usepackage{subfigure}
\usepackage[normalem]{ulem}
\usepackage{relsize}
\usepackage{textcomp}
\usepackage{times}
\usepackage{tikz}
\usepackage{wrapfig}
 \usepackage{xspace}
\usepackage{balance}
\usepackage{thmtools,thm-restate}
\usepackage[normalem]{ulem}
\usepackage{url}
\usepackage{amsmath}
\usepackage[vlined,ruled]{algorithm2e}
\usepackage{xcolor}
\usepackage{mathtools}
\usepackage{xcolor}
\usepackage[utf8]{inputenc}
\usepackage{tabularx}

\usepackage[nocompress]{cite}
\usepackage{hyperref}
\usepackage{cleveref}

\newtheorem{definition}{Definition}

\newcommand{\argmax}{\mathop{\mathrm{argmax}}}

\DeclareMathOperator{\argmin}{argmin}

\begin{document}


\title{Budget-aware Query Tuning: An AutoML Perspective}

\author{
    \alignauthor{
    Wentao Wu
    \hspace{.8cm}
    Chi Wang\\
    \vspace{.2cm}
           Microsoft Research\\
           \vspace{.2cm}
           \{wentao.wu, wang.chi\}@microsoft.com
    }
}

\maketitle

\begin{abstract}
Modern database systems rely on cost-based query optimizers to come up with good execution plans for input queries.
Such query optimizers rely on \emph{cost models} to estimate the costs of candidate query execution plans.
A cost model represents a function from a set of \emph{cost units} to query execution cost, where each cost unit specifies the \emph{unit cost} of executing a certain type of query processing operation (such as table scan or join).
These cost units are traditionally viewed as \emph{constants}, whose values only depend on the platform configuration where the database system runs on top of but are invariant for queries processed by the database system.
In this paper, we challenge this classic view by thinking of these cost units as \emph{variables} instead.
We show that, by varying the cost-unit values one can obtain query plans that significantly outperform the default query plans returned by the query optimizer when viewing the cost units as constants.
We term this cost-unit tuning process ``query tuning'' (QT) and show that it is similar to the well-known hyper-parameter optimization (HPO) problem in AutoML.
As a result, any state-of-the-art HPO technologies can be applied to QT.
We study the QT problem in the context of \emph{anytime tuning}, which is desirable in practice by constraining the total time spent on QT within a given budget---we call this problem \emph{budget-aware} query tuning.
We further extend our study from tuning a single query to tuning a workload with multiple queries, and we call this generalized problem budget-aware workload tuning (WT), which aims for minimizing the execution time of the entire workload.
WT is more challenging as one needs to further prioritize individual query tuning within the given time budget.
We propose solutions to both QT and WT and experimental evaluation using both benchmark and real workloads demonstrates the efficacy of our proposed solutions.
\end{abstract}

\vspace{-0.5em}
\section{Introduction}
\vspace{1em}

\begin{table}
\small
\centering
\begin{tabular}
{|c|r|}
\hline
\textbf{Cost Unit} & \textbf{Default Value} \\
\hline
\emph{seq\_page\_cost} & 1.0\\
\emph{random\_page\_cost} & 4.0\\
\emph{cpu\_tuple\_cost} & 0.01\\
\emph{cpu\_index\_tuple\_cost} & 0.005\\
\emph{cpu\_operator\_cost} & 0.0025\\
\emph{parallel\_tuple\_cost} & 0.1\\
\hline
\end{tabular}
\vspace{-1em}
\caption{Examples of cost units used by PostgreSQL's query planner/optimizer~\cite{pgsql-cost-units}.}
\label{tab:pgsql-cost-units}
\end{table}

Modern database systems rely on cost-based query optimizers to come up with good execution plans for input queries.
Such query optimizers rely on \emph{cost models} to estimate costs of candidate query execution plans.
A cost model represents a function from a set of \emph{cost units} to the query execution cost, where each cost unit specifies the \emph{unit cost} of executing a certain type of query processing operation (such as table scan or join). For instance, Table~\ref{tab:pgsql-cost-units} presents some of the cost units used by PostgreSQL's query planner/optimizer~\cite{pgsql-cost-units}.
This paradigm is followed by other mainstream query processing engines as well, such as Microsoft SQL Server~\cite{sql-server-cost-units}, IBM DB2~\cite{db2-cost-units}, and Apache Spark~\cite{spark-cost-units}.
These cost units are traditionally viewed as \emph{constants}~\cite{pgsql-cost-units}, whose values only depend on the \emph{platform configuration} (e.g., CPU speed) where the query processing engine runs on top of and need to be \emph{calibrated} against the platform~\cite{MackertL86,DuKS92,WuCZTHN13,WuCHN13,WuWHN14,WuNS16}; however, they are \emph{invariant} regardless of the queries being processed by the database system.

In this paper, we challenge this classic view by thinking of these cost units as \emph{variables} instead that can be changed across queries.
We show that, by varying the cost-unit values one can obtain query plans that significantly outperform the default query plans returned by the query optimizer when viewing the cost units as constants.
We call this per-query cost-unit tuning process ``query tuning'' (QT).
At a high-level, query tuning is similar to the \emph{hyper-parameter optimization} (HPO) problem~\cite{HPO-survey} from \emph{automated machine learning} (a.k.a. AutoML), which has received tremendous attention in recent years~\cite{AutoML-survey}.
While the HPO probem aims to find appropriate hyper-parameter values, such as the learning rate and batch size when using stochastic gradient descent (SGD) to train a deep neural network, that can improve the quality of the ML model, QT shares the similar goal of improving the quality of the query plan by seeking appropriate values of the cost units.


The similarity between QT and HPO implies that existing HPO technologies can be directly applied to QT. However, a straightforward application is less attractive, as most of the existing HPO technologies are not \emph{budget-aware}, namely, they do not constrain themselves to conform to a user-specified \emph{tuning time budget}.
There are so far only a few exceptions, including Hyperband~\cite{Hyperband}, BOHB~\cite{BOHB}, CFO~\cite{CFO}, and BlendSearch~\cite{BlendSearch}, which allow user to specify a \emph{timeout} and will exit HPO once the timeout is reached.
However, in practice one often wants to tune multiple queries (called a workload) altogether instead of tuning just one single query.
None of the above work on HPO can be directly applied to this multi-query workload tuning (WT) problem.

In this paper, we study the budget-aware QT and WT problems. We propose solutions to both problems and evaluate their efficacy using both benchmark and real workloads.
The query plans found by our current solutions can significantly improve over the default query plans generated by the query optimizer while conforming to the tuning time budget.

The idea of tuning query performance has been extensively explored in the literature but in different sense compared to the one proposed in this paper.
Lots of recent work has been devoted to the so-called \emph{knob tuning} (see~\cite{knob-tuning-survey} for a recent survey).
Some existing work, e.g., UDO~\cite{UDO}, also views the cost units studied in this paper as tunable knobs, though we think the cost units should be separated from other knobs and are worth its own treatment.
The reason is that many of the knobs are related to the runtime configurations of the database server, e.g., buffer pool size. Changing those knobs often requires a server restart that may not be feasible in many situations (e.g., cloud database services with stringent SLA's).
In contrast, tuning the cost units only affects query optimizer's plan choice, which does not pose significant impact on the runtime state of the database server.
In this spirit, the cost units can be thought of as a certain type of query hint~\cite{power-hint}.
However, the goal of query hint is to constrain the search space of the query optimizer to avoid bad query plans that could have been proposed by the optimizer without such restrictions;
on the other hand, tuning cost units actually gives the optimizer more freedom in terms of proposing query plans, and the decision of picking the best plan is deferred to the moment when actual plan exeuction time is observed.
Due to their overheads, the query/workload tuning technologies studied in this paper can perhaps only be applied to \emph{recurring} queries/workloads, where one can tune the queries/workloads in an offline manner and pay it as a one-time price~\cite{HerodotouB10}.
Nonetheless, given the strong connection between QT and HPO, we believe that more progress can be made in the future to enable QT for tuning more adhoc workloads in an online manner.

\vspace{-0.5em}
\section{Problem Formulation}
\vspace{1em}

We assume that a query optimizer uses a cost model configured with a set of tunable parameters, referred to as \emph{cost units} (ref. Table~\ref{tab:pgsql-cost-units} for examples), to estimate the cost of a candidate query execution plan. The plan with the lowest estimated cost is returned by the optimizer for final query execution.
Without loss of generality, we use $\vec{u}$ to represent the set of tunable cost units as an ordered vector.
Given a query $q$, we use $P(q, \vec{u})$ to represent the query plan returned by the query optimizer with the cost units $\vec{u}$. 
Different values of $\vec{u}$ may therefore yield different query plans. 
We next formulate the problems of budget-aware query tuning and workload tuning.

\vspace{-0.5em}
\subsection{Budget-aware Query Tuning}
\label{sec:problem:qt}
\vspace{1em}

Let $q$ be a specific query, and let $U$ be the search space of $\vec{u}$.
That is, if $\vec{u}=(u_1, ..., u_m)$ where each $u_i$ is within some range/domain $D_i$ (e.g., $u_i\in[a_i, b_i]$), then $U=\prod_{i=1}^m D_i$.
This covers both discrete and continuous cost units, though in practice cost units are typically continuous (within certain ranges).

Let $B$ be a given budget on the tuning time.
Let $\vec{u}_1$, ..., $\vec{u}_K$ be the successive trials on the cost units.
Let $t(q, \vec{u}_j)$ be the execution time of the corresponding query plan $P(q, \vec{u}_j)$, for $1\leq j\leq K$.
The budget constraint can then be expressed as $\sum_{j=1}^K t(q, \vec{u}_j)\leq B$. The problem of budget-aware query tuning is then defined as:

\begin{definition}[Budget-aware Query Tuning]
\label{def:tuning:query-level}
Find $\vec{u}^*=\argmin_{1\leq j\leq K}\{t(q, \vec{u}_j)\}$ w.r.t. the budget constraint 
$$\sum\nolimits_{j=1}^K t(q, \vec{u}_j)\leq B.$$
\end{definition}

\vspace{-1em}
\subsection{Budget-aware Workload Tuning}
\vspace{1em}

Let $W=\{q_1, ..., q_n\}$ be a workload of $n$ queries.
Let $f_i$ be the \emph{frequency} of the query $q_i$ being tuned.
For $q_i$, we define its total tuning time
$$t_i=\sum\nolimits_{j=1}^{f_i}t(q_i, \vec{u}_{ij}),$$
where $\vec{u}_{ij}$ represents (the cost units of) the $j$-th trial of $q_i$.
The budget constraint at workload-level can then be expressed as $\sum_{i=1}^n t_i\leq B$.

\begin{definition}[Budget-aware Workload Tuning]
Find 
$$\vec{u}^*_i=\argmin_{1\leq j\leq f_i}\{t(q_i,\vec{u}_{ij})\}$$ 
for $1\leq i\leq n$ w.r.t. the budget constraint $\sum_{i=1}^n t_i\leq B$.
\end{definition}

\vspace{-1em}
\paragraph*{Remark}
The above definition automatically minimizes the workload execution time w.r.t. the budget constraint on tuning time, which can be expressed as $$t(W, \vec{u}^*_W)=t(q_1,...,q_n, \vec{u}^*_1, ..., \vec{u}^*_n)=\sum\nolimits_{i=1}^n t(q_i, \vec{u}^*_i).$$

\vspace{-1em}
\section{Proposed Solutions}
\vspace{1em}

Below we propose solutions to budget-aware query tuning (QT) and workload tuning (WT).

\vspace{-0.5em}
\subsection{Budget-aware Query Tuning}
\vspace{1em}

Since QT is similar to HPO, in theory any HPO algorithm $\mathcal{A}$ can be adapted to work for QT.
For example, \emph{random search} is a simple but competitive algorithm for HPO~\cite{RandomSearch}. It can be easily customized to a budget-aware algorithm by monitoring the time spent on each random trial and terminating once the timeout is reached.

\vspace{-1em}
\paragraph*{Optimization by Plan Caching} Since different cost units may result in the same query execution plan, it is possible that some plan $P(q,\vec{u}_j)$ is a duplicate of another plan $P(q,\vec{u}_i)$ ($i < j$) in the sequence $\vec{u}_1$, ..., $\vec{u}_K$.
One optimization is therefore to maintain a \emph{cache} for observed plans, if memory is not constrained. If a duplicate plan $P(q,\vec{u})$ is detected, we simply set $t(q,\vec{u})=0$.

\vspace{-1em}
\paragraph*{Optimization by Early Stopping} Since only the optimal $\vec{u}^*$ matters, we can safely stop executing a plan $P(q, \vec{u}_j)$ if $t(q, \vec{u}_j)\geq t(q,\vec{u}^*_j)$, where $t(q,\vec{u}^*_j)$ is the lowest execution time observed up to the $j$-th trial.
In practice, we usually have a default value $\vec{u}_0$ for $\vec{u}$ (e.g., the built-in values such as the ones shown in Table~\ref{tab:pgsql-cost-units} for PostgreSQL).
As a special case of the above ``early stopping'' idea, we can stop executing $P(q, \vec{u}_j)$ if $t(q, \vec{u}_j)\geq t(q,\vec{u}_0)$.
This is also a worst-case scenario, as we have $t(q,\vec{u}^*_j)\leq t(q, \vec{u}_0)$, obviously.

\vspace{-0.5em}
\subsection{Budget-aware Workload Tuning}
\vspace{1em}

WT is more complicated than QT, as one has to decide which query to tune next while conforming to the total budget on tuning time.
We propose the following four strategies: (1) round robin; (2) cost-based prioritization; (3) multi-armed bandit; (4) improvement rate.
Moreover, both the plan-cache based optimization and the early-stopping optimization can be used for WT.

\vspace{-0.5em}
\subsubsection{Round Robin}
\vspace{1em}

The round robin strategy simply rotates among the queries and stops when the tuning time budget is exhausted.
Albeit a simple strategy, it has been deemed as a robust and strong baseline in the literature~\cite{LiZLWZ18}.

\vspace{-0.5em}
\subsubsection{Cost-based Prioritization}
\vspace{1em}

This strategy is inspired by the idea of using a priority queue to prioritize query tuning in Microsoft's Database Tuning Advisor (DTA)~\cite{dta}.
Specifically, we order the queries by their best execution time observed so far and then select the slowest query to tune next.

\vspace{-0.5em}
\subsubsection{Multi-armed Bandit}
\vspace{1em}

This strategy models workload-level tuning as a multi-armed bandit problem. Specifically, we view each query as an arm, and use the well-know UCB1 score~\cite{Auer02,AuerCF02} as the criterion for selecting the next query to tune:
\begin{equation}\label{eq:ucb1}
    \argmax_q \Bigg[\bar{r}(q) + \lambda\cdot\sqrt{\frac{\ln{N}}{f_q}}\Bigg].
\end{equation}
Here, $\lambda$ is a constant that balances \emph{exploration} and \emph{exploitation}. We chose $\lambda=\sqrt{2}$ as suggested in the literature~\cite{uct}.
$f_q$ is the number of times (i.e. frequency) that $q$ is tuned, and $N=\sum_{q\in W} f_q$.

$\bar{r}(q)$ is the \emph{average reward} of $q$.
The reward $r(q, \vec{u})$ of tuning $q$ with cost units $\vec{u}$ is defined as its relative improvement over the query execution time using the default cost units $\vec{u}_0$:
$$r(q,\vec{u})=\max\{1-\frac{t(q, \vec{u})}{t(q, \vec{u}_0)},0\}.$$
That is, we cap the reward at 0 if $\vec{u}$ is even worse than $\vec{u}_0$. This should not occur if the early-stopping optimization is used.
The average reward is therefore
$$\bar{r}(q)=\frac{1}{f_q}\sum\nolimits_{j=1}^{f_q}r(q, \vec{u}_j).$$
Again, we stop when the tuning time budget is exhausted.

\vspace{-0.5em}
\subsubsection{Improvement Rate}
\vspace{1em}

The improvement $I_j(q)$ of a query $q$ is defined as the gap between its best execution time found so far and the default execution time using cost units $\vec{u}_0$. That is,
$$I_j(q)=\max\{t(q, \vec{u}^*_j)-t(q,\vec{u}_0), 0\},$$
for $1\leq j\leq f_q$. Again, we cap the improvement at 0 if $\vec{u}^*_j$ is worse than $\vec{u}_0$, which should not happen if the early-stopping optimization is used.
The improvement rate is then defined as
$R_j(q)=\frac{I_j(q)}{f_q}$.

This strategy always selects the query with the highest improvement rate to tune next, which is inspired by BlendSearch~\cite{BlendSearch}. Again, we stop when the tuning time budget is exhausted.

\begin{figure*}
\centering
\includegraphics[width=\textwidth]{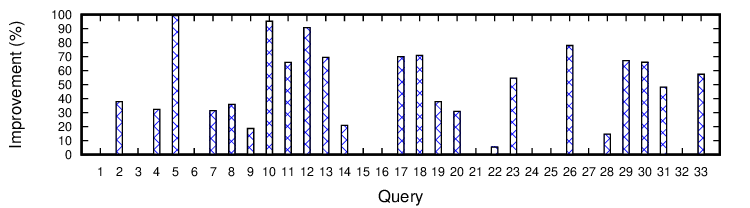}
\vspace{-2.5em}
\caption{Percentage improvement of each query in the \textbf{JOB} workload.}
\label{fig:query-tuning:impr:job}
\vspace{-1em}
\end{figure*}

\begin{figure*}[t]
\centering
\includegraphics[width=\textwidth]{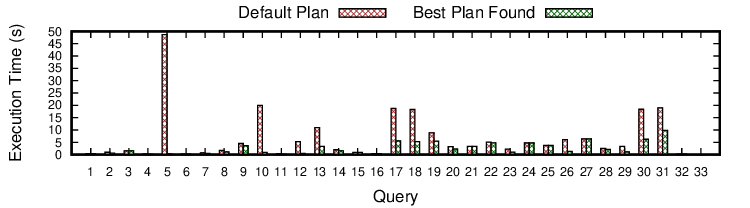}
\vspace{-2.5em}
\caption{Execution time of default plan vs. best plan found by QT for each query of the \textbf{JOB} workload.}
\label{fig:query-tuning:time:job}
\vspace{-1em}
\end{figure*}

\vspace{-0.5em}
\section{Experimental Evaluation}
\label{sec:evaluation}
\vspace{1em}

We report experimental results on evaluating the performance of our proposed budget-aware query and workload tuning technologies.

\vspace{-0.5em}
\paragraph*{Datasets and Workloads}
We used various benchmark and real workloads in our evaluation.
Table~\ref{tab:databases} summarizes the information of the workloads. 
For benchmark workloads, we use the join order benchmark (\textbf{JOB})~\cite{job-queries}, as well as the \textbf{TPC-DS} benchmark with scaling factor 10. \textbf{JOB} contains 113 query instances in total, which are grouped into 33 templates, and we pick one query instance from each template. We use the same protocol for \textbf{TPC-DS}.
We chose these two benchmark workloads due to the diversity and complexity in their queries that offer more opportunities for finding better query plans via query tuning.
We also use two real workloads, denoted by \textbf{Real-A} and \textbf{Real-B}.
Table~\ref{tab:databases} summarizes the key statistics of these databases and workloads. The last two columns represent the average number of joins and table scans contained by a query.

\begin{table}
\small
\centering
\begin{tabular}
{|c|c|c|c|c|c|}
\hline
\textbf{Name} & \textbf{DB Size} & \textbf{Queries} & \textbf{Tables} & \textbf{Joins} 
& \textbf{Scans}\\
\hline
\hline
\textbf{JOB} & 9.2GB & 33 & 21 & 7.9 
& 8.9\\
\textbf{TPC-DS} & \emph{sf}=10 & 99 & 24 & 7.7 
& 8.8\\
\hline
\hline
\textbf{Real-A} & 100GB & 25 & 20 & 6.5 
& 7.2\\
\textbf{Real-B} & 60GB & 16 & 7 & 1.9 
& 2.9\\
\hline
\end{tabular}
\vspace{-0.5em}
\caption{Database and workload statistics.}
\label{tab:databases}
\end{table}

\paragraph*{Experimental Settings}
We perform all experiments using Microsoft SQL Server 2017 under Windows Server 2022, running on a workstation equipped with 2.3 GHz AMD CPUs and 256 GB main memory. We focus on tuning eight cost units that are critical to the costs of workhorse operators such as \emph{table scan}, \emph{index seek}, \emph{sort}, \emph{hash join}, and \emph{nested-loop join}. For each cost unit $c$, we set its range/domain as $[0.1\times c_d, 10\times c_d]$ for exploration (ref. Section~\ref{sec:problem:qt}), where $c_d$ is the default value of $c$. 

\vspace{-0.5em}
\subsection{Query Tuning Results}
\vspace{1em}

We use \emph{percentage improvement} as the performance metric.
Specifically, it is defined as 
$$\text{percentage improvement}=1-\frac{t(P_{\text{best}})}{t(P_{\text{default}})}.$$
Here $t(P_{\text{best}})$ and $t(P_{\text{default}})$ represent the execution time of the best plan found by QT and that of the default plan chosen by the query optimizer (using the built-in values of the cost units), respectively.


For each query, we give the HPO algorithm $\mathcal{A}$ 100 trials, and report the best plan found.
For the HPO algorithm $\mathcal{A}$, we evaluated both \emph{random search}~\cite{RandomSearch} and \emph{SMAC}~\cite{SMAC}, but we found that their performances were similar.
As a result, we only report results by using random search, due to its simplicity and lower overhead.

\begin{table}
\small
\centering
\begin{tabularx}{\columnwidth}{|l|X|X|X|X|}
\hline
\textbf{Name} & \textbf{Default (minutes)} & \textbf{Best Plan (minutes)} & Percentage Improvement  & Tuning Time (minutes) \\
\hline
\hline
\textbf{JOB} & 3.72  & 1.33 & 64.2\% & 130\\
\textbf{TPC-DS} & 4.74 & 3.24 & 31.6\% & 580\\
\hline
\hline
\textbf{Real-A} & 13.96 & 7.36 & 47.3\% & 675\\
\textbf{Real-B} & 13.21 & 8.38 & 36.5\% & 160\\
\hline
\end{tabularx}
\vspace{-1em}
\caption{Summary of query tuning results}
\label{tab:query-tuning}
\end{table}

Table~\ref{tab:query-tuning} summarizes the QT results for the workloads evaluated.
We observe percentage improvement ranging from 31.6\% to 64.2\% at workload level (i.e., the total execution time of all queries).
Figure~\ref{fig:query-tuning:impr:job} further showcases the percentage improvement for each query in the \textbf{JOB} workload, whereas Figure~\ref{fig:query-tuning:time:job} presents the corresponding execution time (in seconds) of the default plan and the best plan found by QT.
Note that we did not test the budget-aware version of QT, as it is a special case of budget-aware WT that will be covered next.

\vspace{-0.5em}
\subsection{Workload Tuning Results}
\vspace{1em}

\begin{figure*}
\centering
\subfigure[\textbf{JOB}]{ \label{fig:workload-tuning:job}
\includegraphics[width=0.23\textwidth]{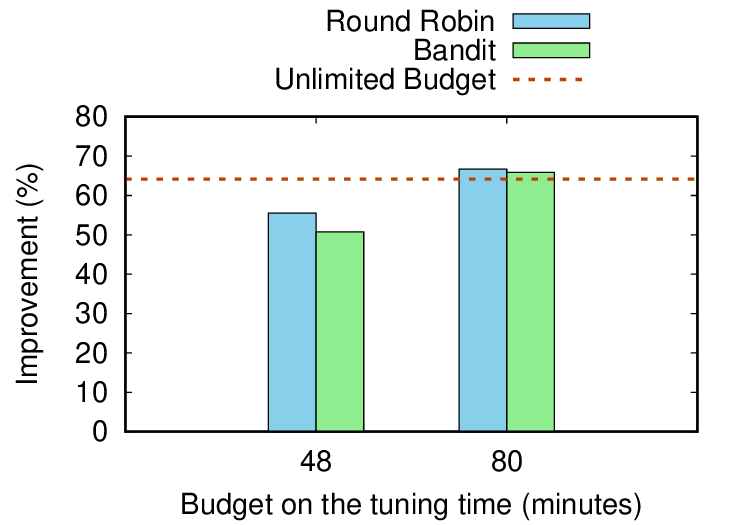}}
\subfigure[\textbf{TPC-DS}]{ \label{fig:workload-tuning:tpcds}
\includegraphics[width=0.23\textwidth]{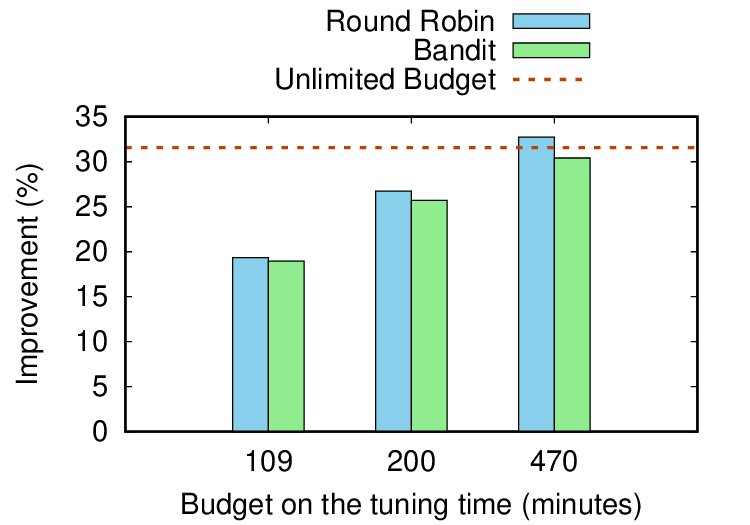}}
\subfigure[\textbf{Real-A}]{ \label{fig:workload-tuning:real-A}
\includegraphics[width=0.23\textwidth]{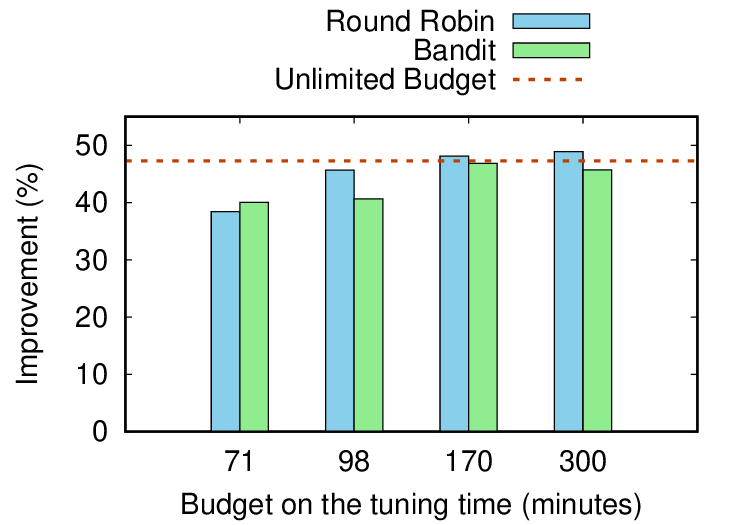}}
\subfigure[\textbf{Real-B}]{ \label{fig:workload-tuning:real-B}
\includegraphics[width=0.23\textwidth]{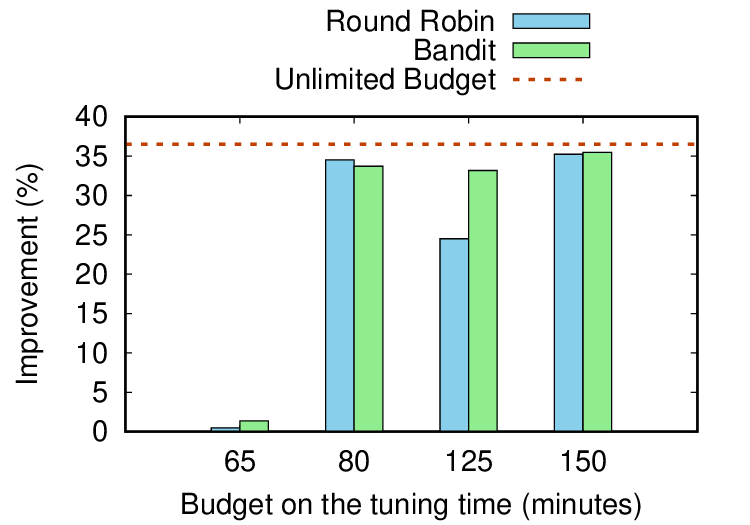}}
\vspace{-1em}
\caption{Percentage improvement resulted from workload tuning when varying the budget on tuning time.}
\label{fig:workload-tuning}
\vspace{-1.5em}
\end{figure*}

\begin{figure}
\centering
\includegraphics[width=0.7\columnwidth]{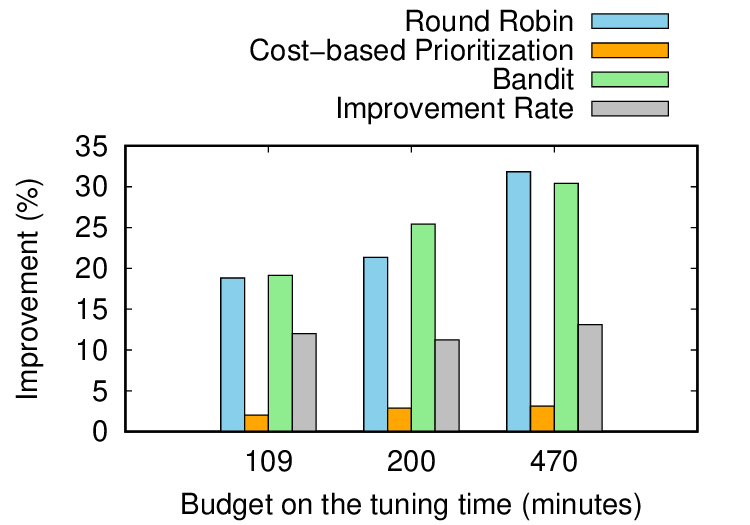}
\vspace{-1em}
\caption{Comparison of WT algorithms on TPC-DS}
\label{fig:wt-algorithms:tpcds}
\end{figure}

We vary the budget on the time given to the workload tuning algorithms and test the percentage improvement at workload-level. Figure~\ref{fig:workload-tuning} presents the results on the workloads tested. The $x$-axis represents the tuning time given to the algorithms, whereas the $y$-axis reports the percentage improvement observed.
To avoid clutter, Figure~\ref{fig:workload-tuning} only includes the two best-performing algorithms, \emph{round robin} and \emph{multi-armed bandit}, which significantly outperform the other two algorithms, \emph{cost-based prioritization} and \emph{improvement rate} (see Figure~\ref{fig:wt-algorithms:tpcds} for a comparison on \textbf{TPC-DS}).
Moreover, the dash line in each chart represents the percentage improvement observed in the QT experiment with 100 trials of random search for each query, whereas the corresponding tuning time has been reported in the last column of Table~\ref{tab:query-tuning}.
We observe that we can obtain similar percentage improvement with much less tuning time.
For example, on \textbf{JOB} it took only 80 minutes for both algorithms to achieve the same percentage improvement, compared to the 130 minutes in QT (i.e., 38.5\% reduction); on \textbf{Real-A}, it took only 98 minutes for \emph{round robin} to achieve the same percentage improvement, in contrast to the 675 minutes taken in QT (i.e., 85.5\% reduction).

\vspace{-0.5em}
\paragraph*{Comparison of Workload Tuning Algorithms}

Figure~\ref{fig:wt-algorithms:tpcds} compares the four workload tuning algorithms with varying tuning time budget, using the \textbf{TPC-DS} workload. We observe that \emph{round robin} and \emph{multi-armed bandit} perform similarly, and they significantly outperform the other two algorithms, \emph{cost-based prioritization} and \emph{improvement rate}.
The \emph{cost-based prioritization} strategy often does not perform well in a budget-aware setting, because it can often get stuck on some very expensive query that is also hard to improve.
On the other hand, the \emph{improvement rate} strategy bypasses this issue by focusing on tuning queries that can improve quickly.
As a result, it improves over the \emph{cost-based prioritization} strategy.
However, it remains less effective compared to \emph{round robin} and \emph{multi-armed bandit}.



\vspace{-1em}
\section{Related Work}
\vspace{-0.5em}
\paragraph*{Autonomous Knob Tuning}
The problem of knob tuning for database systems has attracted intensive research interest (see~\cite{knob-tuning-survey} for a recent survey on this topic).
Some existing work, such as UDO~\cite{UDO}, also views the query optimizer cost units as tunable knobs.
Our view is that the cost units are different from other runtime configuration knobs that require changing the database server's state.
The main impact of the cost units is to influence the query optimizer to generate different query plans.

\vspace{-0.5em}
\paragraph*{Autonomous Index Tuning}
While some existing work also considers index tuning as part of knob tuning (e.g., UDO~\cite{UDO}), we do think that it falls into another special category of database tuning, just like tuning the cost units, which is worth its own treatment.
The reason is similar---index tuning has a different impact on the database system compared to knobs that can change the database server's runtime state.
Specifically, index tuning will change the physical data layout of the database, which in some sense is more dramatic as it has broader impact on various database system components, such as metadata management, query optimizer's plan choice, statistics (such as histograms) maintenance, and so on. We refer the readers to~\cite{ml-powered-index-tuning} for a recent survey on index tuning.
The idea of budget-aware index tuning has also been explored recently~\cite{budget-aware-index-tuning}, with the similar motivation of constraining the tuning time in practice.

\vspace{-0.5em}
\paragraph*{Hyper-parameter Optimization}
The HPO problem has been extensively studied in the literature (see~\cite{HPO-survey} for a survey).
In addition to simple strategies such as \emph{random search}~\cite{RandomSearch} and bandit-based strategies such as Hyperband~\cite{Hyperband}, many HPO strategies rely on classic Bayesian Optimization (BO), such as Hyperopt~\cite{Hyperopt,TPE}, SMAC~\cite{SMAC,SMAC3}, Spearmint~\cite{Spearmint}, BOHB~\cite{BOHB}, and OpenBox~\cite{OpenBox}.
These BO-style HPO strategies view the function from the hyper-parameter values to the ML model quality metric (e.g., accuracy) as a \emph{black box} without utilizing workload properties.
While in this paper we use these off-the-shelf HPO strategies without modification, an interesting future direction to explore is to leverage the similarity among the workload queries~\cite{ISUM} and use that information to improve these BO-style strategies.

\vspace{-0.5em}
\section{Conclusion}
\vspace{1em}

In this paper, we proposed the budget-aware query tuning and workload tuning problems. We highlighted the connection between the query tuning problem and the hyper-parameter optimization (HPO) problem in AutoML, and we proposed solutions based on adapting existing HPO algorithms such as \emph{random search} and \emph{SMAC}.
Experimental evaluation shows that (1) query tuning can result in much faster query plans compared to the ones generated by the query optimizer based on the default values of the cost units; and (2) budget-aware workload tuning using simple strategies such as \emph{round robin} or \emph{multi-armed bandit} can significantly reduce the amount of tuning time at workload-level.

We note that the query and workload tuning technologies proposed in this paper are rudimentary as they are simple applications of existing well-known technologies.
As a result, they should serve as baseline approaches that future research can reference and compare against.
One promising direction for future work, as we briefly mentioned, is to further leverage the similarities among workload queries to improve the BO-style approaches.
For instance, a particular set of cost-unit values may be optimal for multiple queries if they share common SQL expressions.
As a result, one may want to perform a clustering on the queries and tune each group/cluster of queries as an independent, smaller workload.
This can further reduce the tuning time on a large workload, or can have more potential of finding better query plans within a given tuning time budget.
Nonetheless, it also raises new challenges such as how to cluster the queries and how to prioritize among the query clusters during workload tuning, which requires further investigation.

\vspace{-0.5em}
\paragraph*{Acknowledgement}
We thank Anshuman Dutt, Bailu Ding, and Vivek Narasayya for their helpful discussion and valuable feedback on this work.



\vspace{-0.5em}
{\small
\bibliographystyle{abbrv}
\bibliography{./paper}

\begin{thebibliography}{10}

\bibitem{pgsql-cost-units}
Postgresql planner cost constants.
\newblock https://www.postgresql.org/docs/current/runtime-config-query.html,
  2024.

\bibitem{Auer02}
P.~Auer.
\newblock Using confidence bounds for exploitation-exploration trade-offs.
\newblock {\em J. Mach. Learn. Res.}, 3:397--422, 2002.

\bibitem{AuerCF02}
P.~Auer, N.~Cesa{-}Bianchi, and P.~Fischer.
\newblock Finite-time analysis of the multiarmed bandit problem.
\newblock {\em Mach. Learn.}, 47(2-3):235--256, 2002.

\bibitem{Hyperopt}
J.~Bergstra, R.~Bardenet, Y.~Bengio, and B.~K{\'{e}}gl.
\newblock Algorithms for hyper-parameter optimization.
\newblock In {\em {NIPS}}, 2011.

\bibitem{RandomSearch}
J.~Bergstra and Y.~Bengio.
\newblock Random search for hyper-parameter optimization.
\newblock {\em J. Mach. Learn. Res.}, 13:281--305, 2012.

\bibitem{TPE}
J.~Bergstra, D.~Yamins, and D.~D. Cox.
\newblock Making a science of model search: Hyperparameter optimization in
  hundreds of dimensions for vision architectures.
\newblock In {\em {ICML}}, volume~28, pages 115--123, 2013.

\bibitem{power-hint}
N.~Bruno, S.~Chaudhuri, and R.~Ramamurthy.
\newblock Power hints for query optimization.
\newblock pages 469--480, 2009.

\bibitem{sql-server-cost-units}
J.~Chang.
\newblock Sql server query optimizer cost formulas.
\newblock
  https://slidetodoc.com/sql-server-query-optimizer-cost-formulas-joe-chang-3/,
  2010.

\bibitem{dta}
S.~Chaudhuri and V.~Narasayya.
\newblock Anytime algorithm of database tuning advisor for microsoft sql
  server, June 2020.

\bibitem{DuKS92}
W.~Du, R.~Krishnamurthy, and M.~Shan.
\newblock Query optimization in a heterogeneous {DBMS}.
\newblock In {\em {VLDB}}, pages 277--291, 1992.

\bibitem{BOHB}
S.~Falkner, A.~Klein, and F.~Hutter.
\newblock {BOHB:} robust and efficient hyperparameter optimization at scale.
\newblock In {\em {ICML}}, volume~80, pages 1436--1445, 2018.

\bibitem{AutoML-survey}
X.~He, K.~Zhao, and X.~Chu.
\newblock Automl: {A} survey of the state-of-the-art.
\newblock {\em Knowl. Based Syst.}, 212:106622, 2021.

\bibitem{HerodotouB10}
H.~Herodotou and S.~Babu.
\newblock Xplus: {A} sql-tuning-aware query optimizer.
\newblock {\em Proc. {VLDB} Endow.}, 3(1):1149--1160, 2010.

\bibitem{spark-cost-units}
R.~Hu, Z.~Wang, W.~Fan, and S.~Agarwal.
\newblock Cost based optimizer in apache spark 2.2.
\newblock
  https://www.databricks.com/blog/2017/08/31/cost-based-optimizer-in-apache-spark-2-2.html,
  2017.

\bibitem{SMAC}
F.~Hutter, H.~H. Hoos, and K.~Leyton{-}Brown.
\newblock Sequential model-based optimization for general algorithm
  configuration.
\newblock In {\em {LION}}, volume 6683, pages 507--523, 2011.

\bibitem{uct}
L.~Kocsis and C.~Szepesv{\'{a}}ri.
\newblock Bandit based monte-carlo planning.
\newblock In {\em {ECML}}, pages 282--293, 2006.

\bibitem{job-queries}
V.~Leis.
\newblock Join order benchmark.
\newblock https://github.com/gregrahn/join-order-benchmark.

\bibitem{Hyperband}
L.~Li, K.~G. Jamieson, G.~DeSalvo, A.~Rostamizadeh, and A.~Talwalkar.
\newblock Hyperband: {A} novel bandit-based approach to hyperparameter
  optimization.
\newblock {\em J. Mach. Learn. Res.}, 18:185:1--185:52, 2017.

\bibitem{LiZLWZ18}
T.~Li, J.~Zhong, J.~Liu, W.~Wu, and C.~Zhang.
\newblock Ease.ml: Towards multi-tenant resource sharing for machine learning
  workloads.
\newblock {\em Proc. {VLDB} Endow.}, 11(5):607--620, 2018.

\bibitem{OpenBox}
Y.~Li, Y.~Shen, W.~Zhang, Y.~Chen, H.~Jiang, M.~Liu, J.~Jiang, J.~Gao, W.~Wu,
  Z.~Yang, C.~Zhang, and B.~Cui.
\newblock Openbox: {A} generalized black-box optimization service.
\newblock In {\em {KDD}}, pages 3209--3219, 2021.

\bibitem{SMAC3}
M.~Lindauer, K.~Eggensperger, M.~Feurer, A.~Biedenkapp, D.~Deng, C.~Benjamins,
  T.~Ruhkopf, R.~Sass, and F.~Hutter.
\newblock {SMAC3:} {A} versatile bayesian optimization package for
  hyperparameter optimization.
\newblock {\em {JMLR}}, 23, 2022.

\bibitem{db2-cost-units}
G.~M. Lohman.
\newblock The db2 universal database optimizer.
\newblock https://cs.uwaterloo.ca/~ilyas/CS448W14/ibm.pdf, 2003.

\bibitem{MackertL86}
L.~F. Mackert and G.~M. Lohman.
\newblock R* optimizer validation and performance evaluation for distributed
  queries.
\newblock In {\em {VLDB}}, pages 149--159, 1986.

\bibitem{ISUM}
T.~Siddiqui, S.~Jo, W.~Wu, C.~Wang, V.~R. Narasayya, and S.~Chaudhuri.
\newblock {ISUM:} efficiently compressing large and complex workloads for
  scalable index tuning.
\newblock In {\em {SIGMOD}}, pages 660--673. {ACM}, 2022.

\bibitem{ml-powered-index-tuning}
T.~Siddiqui and W.~Wu.
\newblock Ml-powered index tuning: An overview of recent progress and open
  challenges.
\newblock {\em {SIGMOD} Rec.}, 52(4):19--30, 2023.

\bibitem{Spearmint}
J.~Snoek, H.~Larochelle, and R.~P. Adams.
\newblock Practical bayesian optimization of machine learning algorithms.
\newblock In {\em {NIPS}}, 2012.

\bibitem{BlendSearch}
C.~Wang, Q.~Wu, S.~Huang, and A.~Saied.
\newblock Economic hyperparameter optimization with blended search strategy.
\newblock In {\em {ICLR}}, 2021.

\bibitem{UDO}
J.~Wang, I.~Trummer, and D.~Basu.
\newblock {UDO:} universal database optimization using reinforcement learning.
\newblock {\em Proc. {VLDB} Endow.}, 14(13):3402--3414, 2021.

\bibitem{CFO}
Q.~Wu, C.~Wang, and S.~Huang.
\newblock Frugal optimization for cost-related hyperparameters.
\newblock In {\em {AAAI}}, 2021.

\bibitem{WuCHN13}
W.~Wu, Y.~Chi, H.~Hacig{\"u}m{\"u}s, and J.~F. Naughton.
\newblock Towards predicting query execution time for concurrent and dynamic
  database workloads.
\newblock {\em PVLDB}, 6(10):925--936, 2013.

\bibitem{WuCZTHN13}
W.~Wu, Y.~Chi, S.~Zhu, J.~Tatemura, H.~Hacig{\"u}m{\"u}s, and J.~F. Naughton.
\newblock Predicting query execution time: Are optimizer cost models really
  unusable?
\newblock In {\em ICDE}, pages 1081--1092, 2013.

\bibitem{WuNS16}
W.~Wu, J.~F. Naughton, and H.~Singh.
\newblock Sampling-based query re-optimization.
\newblock In F.~{\"{O}}zcan, G.~Koutrika, and S.~Madden, editors, {\em
  {SIGMOD}}, pages 1721--1736. {ACM}, 2016.

\bibitem{budget-aware-index-tuning}
W.~Wu, C.~Wang, T.~Siddiqui, J.~Wang, V.~R. Narasayya, S.~Chaudhuri, and P.~A.
  Bernstein.
\newblock Budget-aware index tuning with reinforcement learning.
\newblock In Z.~G. Ives, A.~Bonifati, and A.~E. Abbadi, editors, {\em
  {SIGMOD}}, pages 1528--1541, 2022.

\bibitem{WuWHN14}
W.~Wu, X.~Wu, H.~Hacig{\"{u}}m{\"{u}}s, and J.~F. Naughton.
\newblock Uncertainty aware query execution time prediction.
\newblock {\em {PVLDB}}, 7(14):1857--1868, 2014.

\bibitem{HPO-survey}
T.~Yu and H.~Zhu.
\newblock Hyper-parameter optimization: {A} review of algorithms and
  applications.
\newblock {\em CoRR}, abs/2003.05689, 2020.

\bibitem{knob-tuning-survey}
X.~Zhao, X.~Zhou, and G.~Li.
\newblock Automatic database knob tuning: {A} survey.
\newblock {\em {IEEE} Trans. Knowl. Data Eng.}, 35(12), 2023.

\end{thebibliography}
}


\end{document}